\documentclass[preprintnumbers,showpacs,twocolumn,amsmath,amssymb]{revtex4}

\usepackage{amsmath}
\usepackage{amsfonts}
\usepackage{amssymb}
\usepackage{verbatim}
\usepackage{graphicx}
\usepackage{float}

\usepackage{color,hyperref}
\usepackage{soul} 

\definecolor{BLACK}{gray}{0}
\definecolor{WHITE}{gray}{1}
\definecolor{RED}{rgb}{1,0,0}
\definecolor{GREEN}{rgb}{0.2,.6,0.2}
\definecolor{BLUE}{rgb}{0,0,1}
\definecolor{CYAN}{cmyk}{1,0,0,0}
\definecolor{MAGENTA}{cmyk}{0,1,0,0}
\definecolor{YELLOW}{cmyk}{0,0,1,0}
\definecolor{GOLD}{rgb}{0.75,0.56,0.00}

\definecolor{ORANGE}{rgb}{0.9,0.3,0}

\newcommand{\ro}[1]{\left( {#1} \right)}

\newcommand{\sch}{Schr{\"o}dinger}

\newcommand{\xfrac}[2]{{#1}/{#2}}
\newcommand{\rfrac}[2]{({#1}/{#2})}

\newcommand{\ket}[1]{| #1 \rangle}
\newcommand{\bra}[1]{\langle #1 |}

\begin{document}

\title{Optimal measurements for tests of Einstein-Podolsky-Rosen-steering with no detection loophole using two-qubit Werner states}

\author{D.\ A.\ Evans and  H.\ M.\ Wiseman}

\affiliation{Centre for Quantum Computation and Communication Technology (Australian Research Council), Griffith University, Brisbane, 4111, Australia\\ and Centre for Quantum Dynamics, Griffith University, Brisbane, 4111, Australia}


\begin{abstract}
It has been shown in earlier works that the vertices of Platonic solids
are good measurement choices for tests of Einstein-Podolsky-Rosen (EPR)-steering using
isotropically entangled pairs of qubits. Such measurements
are regularly spaced, and measurement diversity is a good feature for
making EPR-steering inequalities easier to violate in the presence of
experimental imperfections. However, such measurements
are provably suboptimal. Here, we develop a method for devising optimal
strategies for tests of EPR-steering, in the sense of being most robust 
to mixture and inefficiency (while still closing the detection loophole, of course), 
for a given number $n$ of measurement settings. We allow for arbitrary 
measurement directions, and arbitrary weightings of the outcomes in the 
EPR-steering inequality. This is a difficult optimisation problem for large $n$, 
so we also consider more practical ways of constructing near-optimal EPR-steering 
inequalities in this limit.
\end{abstract}

\keywords{EPR; Steering; detection loophole; inefficiency; loss-tolerance}

\pacs{03.65.Ud, 03.67.Mn, 42.50.Xa}

\maketitle

\section{Introduction}

It is one of the most well-known and unintuitive features of quantum
mechanics that entangled quantum systems can, in a 
way that disturbed Einstein, instantaneously affect each other. 
Specifically, the famous Einstein,
Podolsky, and Rosen (EPR) paper of 1935 \cite{EPR}, which made the first
prediction of this feature, 
used it to argue that quantum mechanics
itself must be incomplete. The EPR paper presents a thought experiment
involving a maximally entangled state of two systems, for which 
measurement of the first (Alice's) system 
 forces the second (Bob's) system into one of a set of basis states, 
with the basis depending on the choice of measurement made upon the first. 
That is, Alice's choice of measurement determines which of Bob's observables is 
predictable by her. But EPR implicitly rule out instantaneous action-at-a-distance, assuming that 
``no real change can take place in the second system in consequence of anything that may 
be done to the first system," (that is, Bob's system is not disturbed \cite{Wis13}, 
explaining why Einstein was). 
Hence they conclude that these different observables must have
well-defined values regardless of Alice's choice of measurement. But quantum mechanics 
forbids simultaneous values for non-commuting observables. Thus, they say,  
``the wave function does not provide a complete description of the physical reality.''

Contrary to EPR, Schr{\"o}dinger argued, in the same year \cite{Schrod}, that quantum mechanics was not incomplete, but idealised. 
He used the term ``steering'' for the effect EPR identified, namely 
that ``as a consequence of two different measurements performed upon the first
system, the second system may be left in states with two different [types of] wavefunctions.''
But he thought this was unrealistic when describing systems that are spatially distant, 
because some sort of decoherence would prevent the entanglement from being established 
in such situations. In this way, he, too, thought that instantaneous 
action {\em at a distance} could be kept out of the most fundamental description of reality.

The EPR paper advocated the possibility of local hidden variables (LHVs) in quantum systems which would account for the illusory (in their view) nonlocality in the theory \cite{locaus,bellcite}.  However, it was proved by Bell in 1964 \cite{Bell} that there exist predictions of quantum mechanics for which no possible LHV model could account. Finally, in 1982, examples of Bell nonlocality were experimentally realised \cite{Aspect}. Even without a loophole-free test of Bell nonlocality, it has become widely accepted that (contrary to \sch's hope) entanglement can exist over long distances, and that Bell nonlocality is real.

Entanglement and Bell nonlocality have been rigorously defined for decades;
however, it was not until relatively recently (2007 \cite{key-4,key-3}) that
the particular class of nonlocality described in the EPR paper was
actually formalised. The ability of an entangled quantum state to
nonlocally affect another (though not necessarily vice versa 
\cite{Olsen,OneWayExp,Brunner}; see also \footnote{We note that the loss-tolerant inequalities discussed below 
from Ref.~\cite{Bennet} also give an example of one-way steering. For Werner states with purity parameter $\mu \in [1/2,1]$, 
if Alice's efficiency is less than $2(1-\mu)$ then EPR-steering by her is impossible, while Bob's efficiency is irrelevant. Motived by this, consider a Werner state of which one side has passed through a lossy channel which 
replaces the qubit state by the vacuum state $\ket{v}$ with probability $p > 2\mu-1>0$. This creates the qutrit-qubit state $(1-p)\left[\mu|\psi_{s}\rangle\langle\psi_{s}|+(1-\mu)I^{\alpha\beta}/4\right] + p \ket{v}\bra{v}\otimes I^\beta/2$, where here $|\psi_{s}\rangle$ is a singlet state in the two-qubit subspace, and 
$I^{\alpha\beta}$ is understood to act only on this subspace. By construction, Alice cannot steer Bob, but Bob can steer Alice because Alice (now considered trusted) can simply consider steering in her qubit subspace.}) has come
to be known as \textit{EPR-steering} \cite{Cavalcanti,Saunders,Bennet,Parsimonious}.

The nonlocality described in the EPR paper 
had been studied mainly in the context of their position--momentum example (see, 
e.g.,~\cite{Reid1989,Reid2009}) but the formal notion introduced in Ref.~\cite{key-4} 
has opened the door to a series of new experiments. Following 
the first demonstration of this general notion of EPR-steering in \cite{Saunders},
 three experiments have each closed the detection loophole in tests of EPR-steering. 
One did so while also closing the locality loophole over 48 m \cite{Vienna} 
(thus definitively disproving \sch's suggested resolution of the EPR paradox). 
Another closed the detection loophole with only two different measurements 
(as in the original EPR scenario) by employing state-of-the-art
transition edge detectors \cite{UQ}. The remaining paper closed
the detection loophole using commonplace photon detectors while also
enduring the losses of transmitting the measured photons through an
extra kilometre of fibre-optic cable \cite{Bennet}.

The accomplishments of this third paper are due to the highly loss-tolerant
EPR-steering criteria that it employed to rigorously close the detection loophole.  
Reference~\cite{Evans} describes the formulation of these criteria in more detail, also 
showing them to be more loss-tolerant than another class of EPR-steering criteria 
(which includes those used in Refs.~\cite{Vienna,UQ}). 
In this paper, we reconsider those criteria and
reveal that they are actually not optimally loss-tolerant EPR-steering
criteria. In doing so, we demonstrate a method for optimising similar
tests of EPR-steering, and show that the optimal measurement strategies
for such an experiment are just as practicable, significantly more more loss-tolerant 
in some regimes, and are (unlike those used in Ref.~\cite{Bennet}) applicable
for an arbitrary number of different measurements by Alice.

In Sec.~II of this paper, we briefly review the operational definition of EPR-steering 
and the family of states we consider in this paper. In Sec.~III we review 
linear EPR-steering criteria, including postselection, then identify and close the inefficient detection loophole 
this potentially incurs \cite{Garg,Cyril}. We then review, in Sec.~IV, the EPR-steering criteria
obtained when using Platonic solid measurement strategies.
We discuss the limitations of Platonic solid strategies, including their inherent restrictions in measurement number $n$ (i.e., $n\leq10$), and consider geodesic solid strategies (introduced for $n=16$ in Ref.~\cite{Bennet}), which circumvent this restriction.

Going from Platonic solids to geodesic solids is a more radical step than it may first appear. 
Because it is no longer the case that every vertex is equivalent to every other, a non-trivial constraint 
can be used to obtain stronger criteria  (than those in Ref.~\cite{Bennet}):  
that, when post-selection by Alice is allowed, the probability of a null result be independent of 
Alice's measurement choice. Moreover, there is no longer any symmetry-based justification
for all vertices to be equally weighted; for a geodesic solid comprising two dual Platonic solids (such as the $n=16$ 
of Ref.~\cite{Bennet}, and $n=7$ here) even tighter criteria will result from weighting the two sets differently. 
All this is introduced in Sec.~IV, and serves as a springboard to the completely general consideration in 
Sec.~V.  There, we allow arbitrary arrangements of $n$ vertices, with arbitrary weighting of each vertex, 
and find still tighter criteria for $n$ ranging from $4$ to $8$. 
For the states we consider, these are the most loss-tolerant EPR-steering criteria
possible for any chosen number of measurements, $n$. We conclude in Sec.~VI 
with a discussion of experimental practicalities and future work.
Therein, we address the benefits and difficulties presented by the most optimal measurement strategies for each $n$, and consider whether optimality alone necessarily makes these the best possible choices for constructing experimental tests of EPR-steering.

\section{Tests Of EPR-steering}

The operational definition of EPR-steering that we employ in this
paper is such that one experimental party, Bob, possesses a quantum
state, and another party, Alice, claims to possess a state that is
entangled with Bob's. Bob asks Alice to make one out of a pre-specified 
set of measurements on her state, and inform him of her results.
Using both Alice's results and the results of his own measurements on his system,
Bob then calculates the value of some EPR-steering parameter
and is only convinced that Alice is telling the truth if there is no local hidden state (LHS) model
which could attain the same value.

LHS models assume that Bob's quantum state is preexisting, and
can only depend on Alice's results as much as can be explained by
some local (to Alice) hidden variable that may be correlated 
with Bob's state. This is used to define EPR-steering bounds
by constructing a theoretical limit on some property of Bob's system,
based on the assumption that Bob's quantum system cannot be nonlocally
affected by Alice's measurements. Thus, a violation of this limit
demonstrates EPR-steering.

The EPR-steering criteria that we will use 
are based upon measurements of qubit observables (typcially photon polarisation, but 
we will also use the terminology of spin). 
Moreover, we specialise to criteria suitable for two-photon entangled states 
that are Werner states: 
\begin{equation}
\rho^{\alpha\beta}=\mu|\psi_{s}\rangle\langle\psi_{s}|+(1-\mu)\frac{\mathbb{I}^{\alpha\beta}}{4},\label{eq:Werner}
\end{equation}
where $|\psi_{s}\rangle$ represents the spin singlet state: $|\psi_{s}\rangle=\rfrac{1}{\sqrt{2}}(|0\rangle^{\alpha}\otimes|1\rangle^{\beta}-|1\rangle^{\alpha}\otimes|0\rangle^{\beta})$.
The $\alpha$ and $\beta$ superscripts respectively denote properties
of Alice's and Bob's subsystems. The second term represents
pairs of qubits that are uncorrelated, and the first term represents
qubits that are maximally entangled. Thus the purity parameter $\mu \leq 1$
determines the degree of entanglement in the ensemble $\rho^{\alpha\beta}$,
with entanglement being present for $\mu>1/3$ \cite{Werner}.

\section{Linear Convex Criteria}

We will consider EPR-steering criteria that are analogous to (linear) entanglement 
witnesses \cite{HorEtalPLA96}. That is, the expectation value
of a correlation function between Alice and Bob's spin measurements,
summed over the measurement settings. Since in tests of EPR-steering 
we cannot trust Alice's detectors or the results she states \cite{key-4,key-3}, this 
correlation function must be defined generally as a classical expectation value 
over Alice's reported result $A_r$, denoted by $E_{A_{r}}$, as follows: 
\begin{eqnarray}
S_{n}&=&-\frac{1}{n}\sum_{r=1}^{n}E_{A_{r}}\left[A_{r}\langle\hat{\sigma}_{r}^{\beta}\rangle_{A_{r}}\right], 
\label{eq:Sn}
\end{eqnarray}
where each $r$ denotes a particular measurement setting on the Bloch sphere,
and $n$ denotes the total number of such settings. Bob's qubit observable 
is $\hat{\sigma}_{r}^{\beta}$, and $A_{r} \in \{-1,1\}$ 
is the result Alice submits for her measurement. We can restrict 
Alice's results to these values of equal magnitude 
because of the symmetry of the Werner state.

If Alice, and her detectors, 
were trustworthy, then the result $A_r$ would correspond to a measurement of 
her qubit observable $\hat{\sigma}_{r}^{\alpha}$. Then the 
correlation function between Alice and Bob's results can be written as
\begin{eqnarray}
S_n &=& -
\frac{1}{n}\sum_{r=1}^{n}\sum_{A_{r}}P(A_{r})A_{r}\langle\hat{\sigma}_{r}^{\beta}\rangle_{\rho_{A_{r}}^{\beta}}
\\ 
&=&-\frac{1}{n}\sum_{r=1}^{n}\langle\hat{\sigma}_{r}^{\alpha}\hat{\sigma}_{r}^{\beta}\rangle, 
 \end{eqnarray} 
where $\rho_{A_r}^{\beta}$ is the state of Bob's system, conditioned upon $A_r$ being the result of Alice's measurement.
If Alice and Bob share an entangled state as in Eq.~(\ref{eq:Werner}), and $\hat{\sigma}_{r}^{\alpha}=\hat{\sigma}_{r}^{\beta}$, then
the value of this function is easily shown to be $\mu$.

However, Bob must consider that Alice might not share an entangled state with him, and could be employing an LHS model,
in which case $S_{n}$ would be calculated from
\begin{equation}
S_{n}=-\frac{1}{n}\sum_{\xi}P(\xi)\sum_{r=1}^{n}A_{r,\xi}\langle\hat{\sigma}_{r}^{\beta}\rangle_{\rho_{\xi}^{\beta}},\label{eq:LHS}
\end{equation}
where $\xi$ represents the local hidden variable(s) inherent to Bob's
system, upon which Alice bases her knowledge of Bob's state. In this scenario, Bob receives
each state $\rho_{\xi}^{\beta}$ with probability $P(\xi)$, and Alice
submits results $A_{r,\xi}$ dependent upon both $r$ and $\xi$.
This expression relies on the assumption that there is an LHS model
of Bob's system, the existence of which means that there is a bound upon Eq. (\ref{eq:LHS})
that is not present in a quantum mechanical system \cite{Cavalcanti}.

In order to ensure that this is as rigorous a test as possible, in
defining our EPR-steering bound we will assume that Alice controls 
anything that depends upon the hidden variable(s),
$\xi$; namely, $P(\xi)$, $\rho_{\xi}^{\beta}$, and $A_{r,\xi}$. 
Note that the only thing that does not have any dependence upon $\xi$
is Bob's choice of measurement. 
The assumption of locality in this
LHS model is manifested in Alice's inability to influence or predict
Bob's measurement choice. To this end, we must assume that Bob randomises
the order in which he performs each of his measurements, and that
Alice does not have foreknowledge of, or access to, his random number
generation (this is referred to in other works as the Free Will Assumption
\cite{Bell}, which we will not be further addressing). 

Under the above conditions, it is apparent that $-\sum_{r}A_{r,\xi}\langle\hat{\sigma}_{r}^{\beta}\rangle_{\rho_{\xi}^{\beta}}$ 
is bounded above by $\sum_{r}|\langle\hat{\sigma}_{r}^{\beta}\rangle_{\rho_{\xi}^{\beta}}|$, 
which is always achievable by choosing a suitable sign for $A_{r,\xi}$. 
A proof of this, and of which ensembles of states a cheating Alice can use to attain this optimal value, are given in Ref.~\cite{Evans}.
But if the only concern is to maximise $S_n$ (an assumption to which we will return in Sec.~\ref{sec:globaloptimal}) 
then this can clearly be achieved for a single state $\rho_{\xi}^{\beta}$.

Even if there were more than one state that maximised $S_n$, there is no reason (at this stage) for Alice to use more than one. 
Therefore, we can take $P(\xi)=1$ for that state, and $\xi$ will now denote any choice that maximises $S_{n}$.
The $A_{r,\xi}$ values corresponding to this choice are obviously 
$A_{r,\xi}=- \textrm{sign}(\langle\hat{\sigma}_{r}^{\beta}\rangle_{\rho_{\xi}^{\beta}})$.
However, to evaluate the bound on $S_n$ it is more convenient to keep $A_{r}$, 
writing
\[
S_{n}=-\frac{1}{n}\sum_{r=1}^{n}A_{r,\xi}\langle\hat{\sigma}_{r}^{\beta}\rangle_{\rho_{\xi}^{\beta}}=-\left\langle \frac{1}{n}\sum_{r=1}^{n}A_{r,\xi}\hat{\sigma}_{r}^{\beta}\right\rangle _{\rho_{\xi}^{\beta}},
\]
with the representation on the right being included to highlight that
this entire value can be considered as the expectation value of an operator.  
To seek out the largest possible
value of this expression, we will use the fact that the largest possible
expectation value of any operator is equal to the largest eigenvalue
of that operator. Therefore, the EPR-steering bound we can derive
for $S_{n}$ is 
\begin{equation}
S_{n}\leq k_{n}\equiv\underset{\{A_{r}\}}{\max}\left[\lambda_{\max}\left(\frac{1}{n}\sum_{r=1}^{n}A_{r}\hat{\sigma}_{r}^{\beta}\right)\right],\label{eq:kbound}
\end{equation}
where $\lambda_{\max}$ denotes the maximum eigenvalue of this operator, and the other
maximisation is over the $n$ values of $A_{r}$. 

It should be noted that the normalisation factor of $1/n$ 
in all of the above expressions, stemming from its introduction in Eq.~(\ref{eq:Sn}), is generally paired with the sum over $n$ measurements so that the values of $S_n$ (and related quantities) are limited to $-1\leq S_n \leq 1$.
This restricts the values of $S_n$ to the same range for any $n$-value, allowing meaningful comparison between them.
While it seems logical to weight each measurement result equally, by applying $\xfrac{1}{n}$ to each term or to the whole sum, we will re-evaluate this assumption in Sec.~IV.C.


\subsection{The Inefficient Detection Loophole}

In keeping with our assumption of locality, any null results that
Bob obtains for his measurements cannot be predicted by, or used to
any advantage by a cheating Alice in an LHS model. Because we trust
that Bob's state, and his measurement thereof, is governed by our quantum
mechanical model of it, we can assume that Bob's probability of missing
any result is independent of the value that result would have taken
(had it not been null). Therefore, we will assume that the probability
distribution of the results Bob did not obtain would have been the
same as the probability distribution of Bob's measured results.

This is known as a fair sampling assumption (FSA), and is generally
valid for quantum systems as it is based upon the principles of quantum
mechanics (in the behaviour of detectors). However, since we cannot
assume that Alice's results are generated through measurement of a
quantum state, we cannot apply any FSA to her results in any test
of EPR-steering (which is, in part, a test of quantum mechanics itself).
To simply postselect out any of Alice's null results would open an
inefficient detection loophole in our test.

\subsection{Inequalities Allowing Post-selection}

Even though the FSA cannot be made for Alice, this does not mean 
that it is not permissible to postselect on Alice getting (or claiming to get) 
a non-null result. This postselection is permissible as long as the 
bound $k_n$ in the inequality Eq.~(\ref{eq:kbound}) is adjusted (to a higher value, naturally), 
to take into account the extra flexibility offered to a dishonest Alice if she 
is allowed to submit null results with a certain probability $1-\epsilon$. 
Since Bob has no way of knowing whether this probability is due to genuine
inefficiencies or not, we refer to $\epsilon$ (such that $0\leq\epsilon\leq1$), 
as Alice's \textit{apparent efficiency}. Alice's optimal cheating strategies, 
which gives us the new bounds $k_n(\epsilon)$ for the post-selected 
correlation function, were derived in Ref.~\cite{Bennet}, with more 
details in Ref.~\cite{Evans}. The analysis in the remainder of the 
present paper builds on this, so we briefly review it here.

If Alice chooses to submit non-null results only for a predetermined set 
of $m$ measurement settings, with $m\leq n$, 
her optimal $\rho_{\xi}^{\beta}$ is defined
by the values of these $m$ settings. Such a strategy can be referred
to as a \textit{deterministic} strategy, and the maximal $S_{n}$ values
obtainable with such a strategy are calculated to be
\begin{equation}
D_{n}(\epsilon_{m})=\underset{\{A_{r}\}_{\epsilon_{m}}}{\max}\left[\lambda_{\max}\left(\frac{1}{n}\sum_{r}A_{r}\hat{\sigma}_{r}^{\beta}\right)\right],
\label{eq:platdet}
\end{equation}
where $\epsilon_{m}=\xfrac{m}{n}$ is the apparent efficiency associated
with any such strategy, which is necessarily constrained to be $\epsilon_{m}\in\{\xfrac{1}{n},\xfrac{2}{n},\ldots,\xfrac{(n-1)}{n},1\}$.
The sum in the above expression can be over either $n$ or $m$ settings,
since the maximisation over $\{A_{r}\}$ is constrained such that
a portion $\epsilon_{m}n=m$ of the $A_{r}$ values will be nonzero.

An experimental determination of $S_{n}$ would require many
repetitions for each of the $n$ settings, and Alice is not constrained
to choose the same measurements to be null in every iteration, nor
even to choose the same number of nulls in every iteration. If Alice
uses a combination of deterministic strategies---a \textit{nondeterministic}
strategy---she is also able to avoid constraining her apparent efficiency
to be $\epsilon\in\{\epsilon_{m}\}$. If using a nondeterministic
strategy, the maximal $S_{n}$ value attainable for any apparent efficiency
$\epsilon$ is
\begin{equation}
K_{n}(\epsilon)=\underset{\{w_{m}\}}{\max}\left[\sum_{m=1}^{n}w_{m}D_{n}(\epsilon_{m})\right],\label{eq:NDbound}
\end{equation}
where $w_{m}$ defines the weighting with which Alice uses each deterministic
strategy, each of which is defined by its apparent efficiency, $\epsilon_{m}$. Thus, the
sum over $m$ indexes all optimal deterministic strategies Alice could
use (there is no benefit for Alice to ever use suboptimal deterministic
strategies, so they are not considered). The weightings $w_{m}$ are
normalised by $\sum_{m}^{n}w_{m}=1$, and constrained such that $\sum_{m=1}^{n}w_{m}\epsilon_{m}=\epsilon$.
It can be seen from the form of Eq. (\ref{eq:NDbound}) that $K_{n}(\epsilon_{m})\geq D_{n}(\epsilon_{m}) \ \forall\epsilon_{m}$.

\label{PlatBound} 

The above construction gives the bound a dishonest Alice can achieve 
for the non-postselected correlation function. Since she declares non-null 
results with probability $\epsilon$ (which is a quantity Bob directly 
calculates from the statistics of her declared results), the bound on the post-selected 
function $S_n$ will be 
\begin{equation} \label{knep}
k_{n}(\epsilon)=\frac{1}{\epsilon}K_{n}(\epsilon).
\end{equation}
%
%

\section{Bob's Measurement Strategies}

Linear EPR-steering criteria of the above form have been studied  
before, both with the FSA for Alice \cite{Saunders} 
and without (i.e., closing the detection loophole) \cite{Bennet,Evans}. 
In all of these works, measurement orientations that are regularly spaced about 
the Bloch sphere were used. 
That is, the spacing between vertices is the same for any pair of nearest neighbours.
The only such arrangements that exist are those
with 2, 3, 4, 6, or 10 different measurement axes, which correspond
to the vertices of the three-dimensional Platonic solids (with the exception
of $n=2$ for which the tetrahedron, whose vertices do not come in 
antipodal pairs, was replaced by the square). Regularly spaced measurements are as far apart as it
is possible to be from their nearest neighbours on the Bloch sphere, 
and in this sense are as different as possible. This minimises
the ability of Alice to choose a state $\rho_{\xi}^{\beta}$ that
leads to high values of $\langle\hat{\sigma}_{r}^{\beta}\rangle_{\rho_{\xi}^{\beta}}$
for many $\hat{\sigma}_{r}^{\beta}$. Intuitively, this seems like
a good choice for making it as hard as possible for a cheating Alice
to obtain high $S_{n}$ values, thereby making the rigorous EPR-steering
bounds as low as possible, and thus making it as easy as possible
for an honest Alice to violate the bound.
It should be noted that this reasoning would not necessarily apply for all kinds of photon polarisation states, 
as it relies on the symmetry of Werner states, which are invariant under 
identical unitary transformations performed on both sides.

\begin{figure}
\includegraphics[width=.9\linewidth]{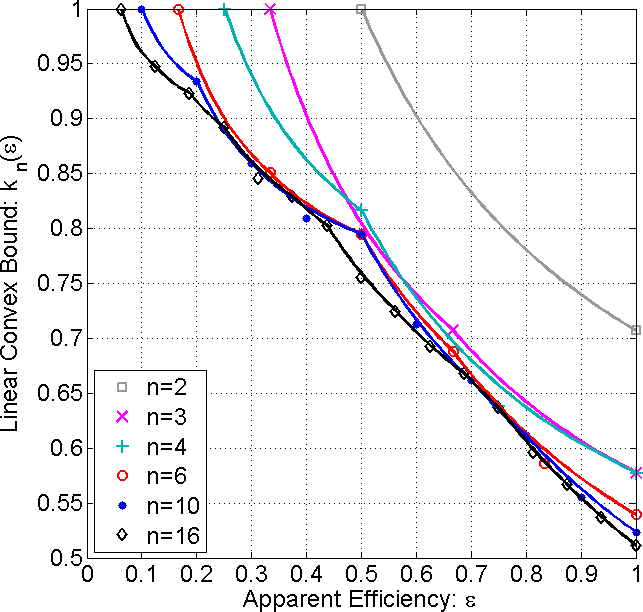}
\caption{Deterministic (the points) and nondeterministic (the lines) postselected
bounds on $S_{n}$, using Platonic solid measurements (and geodesic measurements, for $n=16$).}
\label{fig:Platonic}
\end{figure}

Figure \ref{fig:Platonic} displays the EPR-steering bounds calculated from Eq.~(\ref{knep}) with measurement
orientations defined by Platonic solid vertices. 
Looking closely at this graph, one can observe
that the Platonic solid measurements for $n=4$ are clearly not optimal in general, 
since they give a bound above that for $n=3$ for $0.48\lesssim\epsilon\lesssim0.58$. 
An optimal set of four measurements would never require a higher degree of correlation to demonstrate 
EPR-steering than any set of three measurements. We will return to this issue in Sec.~\ref{sec:globaloptimal}.

Recall that for a Werner state the degree of post-selected correlation $S_n$ is $\mu$, 
which can approach unity. Thus we see that for $\mu$ close to one, the bounds 
$k_n(\epsilon)$ are quite loss-tolerant, especially as
$n$ increases. Indeed, if $\mu=1$, EPR-steering is demonstrable
so long as $\epsilon>\xfrac{1}{n}$. Moreover, in almost all places,
use of more measurements results in EPR-steering bounds that are more
loss-tolerant.

However, regularly spaced measurementsets do not exist for any $n$ above 10,
so we must abandon our scheme of using
regularly spaced measurements if we wish to use $n>10$. But on the
other hand, our restriction to regularly spaced measurements was based
upon the intuition that they were the best choice for their respective
numbers of measurements,   
whereas this is demonstrably not true everywhere, as discussed above.
Therefore, there may be little reason to continue imposing this condition, and little reason to thusly limit our measurement number.

\subsection{Geodesic Solids}

The reader may notice that Fig.~\ref{fig:Platonic} includes not only the Platonic solid
bounds mentioned above, but also includes a bound for $n=16$ measurements, which cannot correspond to any 
Platonic solid. This was derived, and employed experimentally, 
in Ref.~\cite{Bennet}. The measurement orientations used to obtain
this bound correspond to the vertices of a shape that incorporates
the vertices of the icosahedron ($n=6$) and the dodecahedron ($n=10$),
face-centred on one another (as these two shapes are a dual pair).
The resulting arrangement of vertices creates a shape that is a geodesic
solid---each face is an isoceles triangle, so its neighbouring vertices are not
regularly spaced, but are quite close to it. This characteristic is
true of any geodesic solid, so given the obvious benefits of using
this $n=16$ arrangement, it would seem that geodesic solids are
one possible solution for obtaining high-$n$ measurement sets with
robust bounds. Construction of a geodesic solid does not require
two Platonic solids to be superimposed, but only requires vertices
to be added to the face centres of a Platonic solid, or another geodesic
solid. Thus, they cannot be constructed with arbitrary numbers of
vertices, but there does not exist any upper bound upon the number
of vertices that can be used to construct one.

 Having seen that the Platonic solids are not necessarily optimal anyway,
the fact that the vertices of a geodesic solid are not regularly spaced is not really
much of a drawback. Indeed, the viability of geodesic solids may even raise
the point of whether a little asymmetry may be more optimal than regularly
spaced measurements even for small $n$. This will be fully explored in Sec.~\ref{sec:globaloptimal}. 
Meanwhile, we will use the geodesic solids as a first investigation into the 
way asymmetry can affect the derivation of EPR-steering bounds, enabling 
more loss-tolerant tests than any previously calculated.

\subsection{Measurement-independent null result rates}


When a cheating Alice suspects that Bob is keeping track of her null result distribution, her foremost consideration in optimising $S_n$ will be to ensure that this distribution reflects the same profile as that of an honest Alice. This means that Alice should ensure the probability of her
reporting a null result on any given measurement is equal to the probability
of her reporting a null result on any other measurement.
She must do this, even if submitting nulls more often for some measurements
would allow her to obtain a higher $S_{n}$ value. 
In other words, if Bob does verify that the null rate is independent of Alice's supposed setting, 
then he will be convinced of the reality of EPR-steering for a lower $S_n$ value than 
without this verification, thereby making the test more loss-tolerant. 

The uniform spacing of the Platonic solids' vertices grants them large symmetry groups;
the group of all transformations 
which leave the polyhedron invariant. In particular, all vertices are equivalent under the action of each solid's symmetry group.
Therefore any cheating strategy Alice adopts performs precisely as well if it is symmetrised 
by application of the symmetry group, and this ensures that the null-rate can be made independent of Alice's supposed setting.
For example, when $m=2$ for any of the Platonic measurement sets, Alice's optimal choice of $\rho_\xi^\beta$ is any state with its spin axis centred on an edge of the Platonic solid (i.e., equidistant between any pair of adjacent vertices). Such a strategy is equally optimal regardless of which adjacent vertex pair is chosen because all edges are the same length. 

But for any geodesic solid, 
not all edges are the same length, so (considering $m=2$ again) not all edge-centres correspond to optimal strategies. Thus, it may not necessarily be possible to use a nondeterministic strategy that both attains the maximal $S_n$ value \textit{and} keeps Alice's null probabilities equal.
Such limitations would be expected to become even more important for the more complicated cheating strategies [which would be strategies near
$\epsilon = \xfrac{(1+n)}{2n}$: the middle of each curve].

\begin{figure}
\includegraphics[width=.9\linewidth]{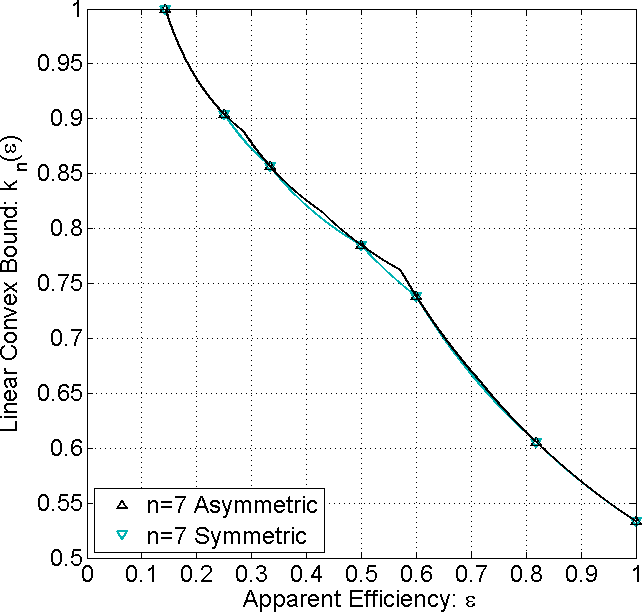}
\caption{Postselected bounds for $n=7,$ with and without consideration given to the symmetry condition.}
\label{fig:n7}
\end{figure}

\begin{figure*}
\includegraphics[width=0.9\linewidth]{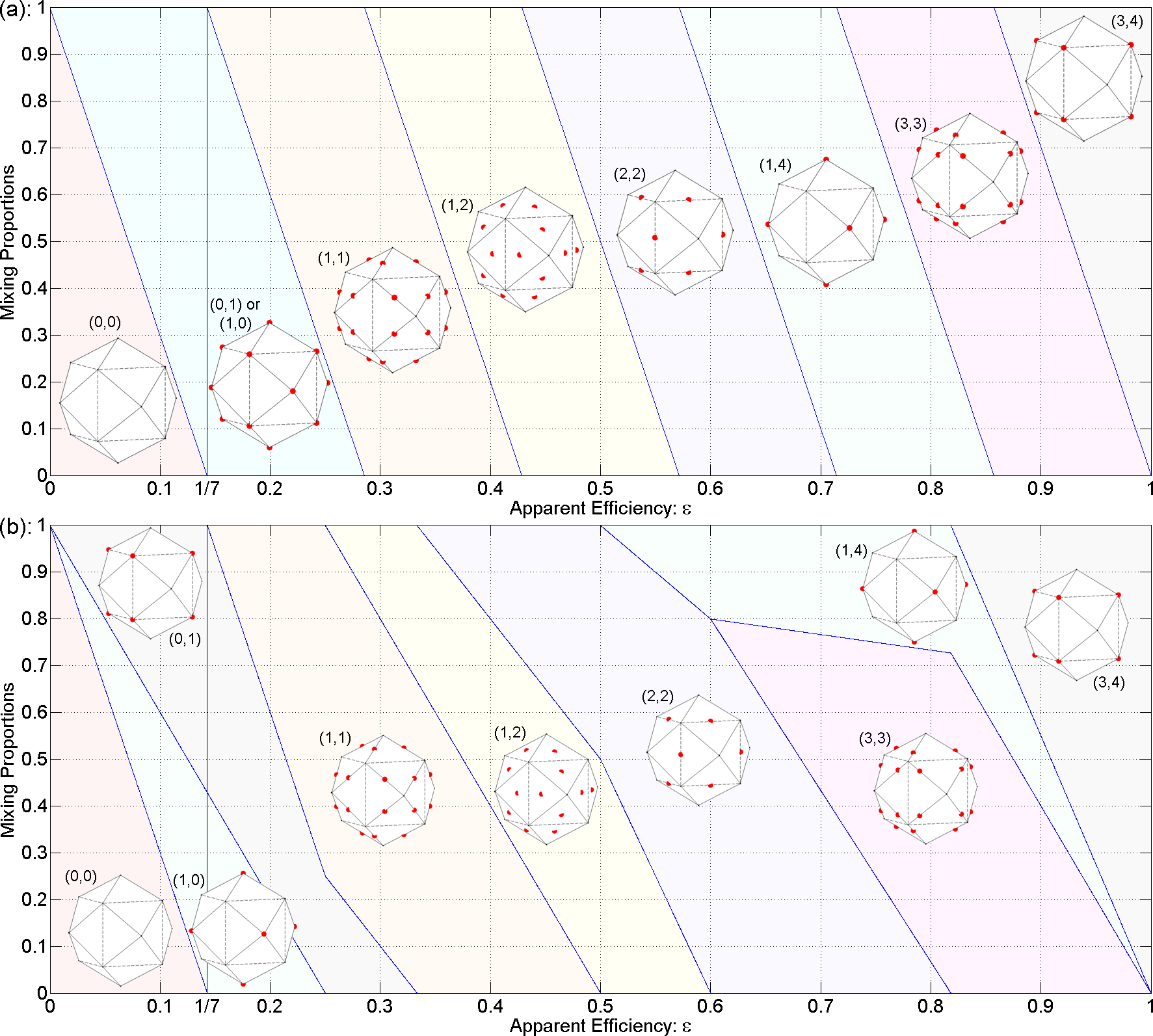}
\caption{The optimal mixing proportions of Alice's deterministic cheating strategies when she is maximising $S_{n}$ for $n=7$ without (a), and with (b) the symmetry constraint upon $A_{r}$.}
\label{fig:weighting7}
\end{figure*}

For illustration, let us consider the geodesic solid that is constructed
by combining the $n=3$ and $n=4$ Platonic solid vertices (because they are dual to one another), 
to obtain $n=7$. To simply maximise the numerical value of $S_n(\epsilon)$, without constraining
her null probabilities for each measurement to be equal, Alice can obtain the ``asymmetric''
bound in Fig.~\ref{fig:n7}. If a cheating Alice takes care
to obey this symmetry condition, then the maximum $S_{n}(\epsilon)$
she can attain is the ``symmetric'' bound in Fig.~\ref{fig:n7}. 
The difference is negligible for most efficiencies,
and is most significant near $\epsilon=1/2$. A
clearer plot of the numerical difference between these two bounds is shown later.


Figure \ref{fig:weighting7} shows how a cheating Alice must depart from her reasonably simple asymmetric
strategy in order to attain the maximum bound under the symmetry condition. The partitions
in this figure show the optimal mixture of deterministic strategies by Alice, 
for each possible $\epsilon$-value. The 
height of each partition
represents the weighting with which Alice must send Bob
each of the ensembles displayed on the shape within that section, in order
to attain the maximal value of $S_{n}$. For example, Alice's optimal
symmetric strategy for $\epsilon=0.3$ requires Alice to choose Bob's
states $\rho_{\xi}^{\beta}$ such that: $10\%$ come from the
ensemble shown on the solid labelled (0,1), $70\%$
from the (1,1) solid, and $20\%$ from the (1,2) solid. The 
states in each ensemble must also be submitted equally frequently, e.g., in this strategy, the eight states on the (0,1) solid must each be submitted ${10\%}/{8}=1.25\%$
of the time, in total. 

The bracketed numbers $(m_{3},m_{4})$ that label each solid in Fig.~\ref{fig:weighting7}
respectively represent the number of non-null responses, for the associated deterministic strategy, 
to Bob's $n=3$ and $n=4$ measurements (that make up the $n=7$ set). 
For a  deterministic strategy $i$, identified with the  pair
$(m^{i}_{3},m^{i}_{4})$, we calculate
the deterministic bound quite similarly to before, as 
\begin{equation}
D_{n}(i)=\underset{\{A^i_{r}\}}{\max}\left[\frac{1}{n} \lambda_{\max} \left(\sum_{r=1}^{n_3} A^i_{r}\hat{\sigma}_{r}^{\beta}+\sum_{r=n_4}^{n_3+n_4} A^i_{r}\hat{\sigma}_{r}^{\beta}\right)\right], 
\label{eq:dn34}
\end{equation}
where $\hat{\sigma}_{r}^{\beta}$ corresponds to Bob's $n=3$ measurements for the $1\leq r \leq n_3\equiv 3$ (the first sum), and for the $n=4$ measurements for $n_4 \equiv 4\leq r \leq 7 = n_3+n_4$ (the second sum).
The index $i$ is over all possible combinations of $(m^{i}_{3},m^{i}_{4})$ and thus 
the maximisation considers the optimal deterministic ensembles for every such combination
[there will be $(n_{3}+1)(n_{4}+1)$ of these]. 
An optimal nondeterministic strategy is 
composed of these $D_n(i)$ as 
\begin{equation}
K_{n}(\epsilon)=\underset{\{w_{i}\}}{\max}\left[\sum_{i}^{(n_{3}+1)(n_{4}+1)}w_{i}D_{n}(i)\right],
\label{eq:kngeo}
\end{equation}
where 
$w_i$ is the weighting of each deterministic strategy, $D_n(i)$, and is constrained such that $\sum_i w_i =1$, 
and such that the apparent efficiency of the strategy is 
$\epsilon=\sum_{i}^{(n_{3}+1)(n_{4}+1)}w_{i}\ro{m^{i}_{3}+m^{i}_{4}}/\ro{n_{3}+n_{4}}$. 
Although constructed slightly differently, these are the same
relations as given in Sec.~\ref{PlatBound}.
In order for $K_n (\epsilon)$ to give the optimal {\em symmetric} nondeterministic bound, we must also constrain
Alice's null probability to be independent of Bob's measurement orientation.
This can be done by constraining $w_i$ such that the mixing of strategies must be in proportions where, over the entire nondeterministic strategy, the null probability for $n=3$ is equal to that for $n=4$. Therefore, $w_i$ must also satisfy
\begin{equation}
\sum_{i}^{(n_{3}+1)(n_{4}+1)}w_{i}\frac{m^{i}_{x}}{n_{x}}=\epsilon,
\label{eq:sym7}
\end{equation}
for both $x=3$ and $x=4$.
%
Without this constraint, the optimal cheating strategies for Alice [those shown in Fig.~\ref{fig:weighting7}(a)]
would lead to very asymmetric reporting of null results. 
For example, at $\epsilon=\xfrac{5}{7}\approx 0.714$ there is a single deterministic strategy; the $(1,4)$ strategy, with an apparent efficiency of $\epsilon=\xfrac{5}{7}$. This strategy requires $\epsilon_{3}=\xfrac{1}{3}$ and $\epsilon_{4}=1$, which means Alice would never report a null result for one of Bob's measurements drawn from the cube ($n=4)$ but would report a null result $2/3$ of the time for one drawn from the octahedron ($n=3$).

\subsection{Weighting for the different types of vertices} \label{sec:IVC}

We have seen that for $n=7$, a cheating Alice is
able to attain a symmetric bound almost always as high as her asymmetric
bound, but only if she employs more elaborate mixings of her deterministic strategies. 
Indeed, at almost every $\epsilon$-value, the optimal symmetric mixings include more deterministic strategies than just the strategies used to attain the optimal asymmetric bounds at that $\epsilon$-value.
Clearly, only when using two (or more) geometrically inequivalent subsets of measurement direction (as in geodesic solids) could
any cheating strategy attain a higher $S_n(\epsilon)$ with an asymmetric null distribution than is possible with a symmetric null distribution.
Thus, only when using such inequivalent measurement subsets can a symmetry condition be used to improve our EPR-steering bounds (as we observed for $n=7$).
%

From this observation, one may come to suspect a further advantage that may be gained in this situation, as follows.
Say an optimal asymmetric cheating strategy involves Alice reporting more null results for  
one of the measurement sets (e.g., the $n_3$ set). This suggests that a cheating Alice would
prefer not to have to report outcomes for this set at all. Therefore,
if Alice were not only forced to report results for these measurements equally often, but actually 
\textit{more} often than other measurements, this would, intuitively, make it harder for a cheating 
Alice to achieve a high correlation $S_n$, averaged over all reported results, especially when we impose the restriction that the cheating 
strategy be symmetric. Thus, using different weights for different measurement sets in the expression for the 
EPR-steering correlation function could conceivably lower our EPR-steering bounds even further.
 
Like the symmetry condition, such an advantage would clearly only be available to Bob if the set of measurements he employs are not regularly spaced. 
To make use of this, we should recall that each measurement was equally weighted in all of our previous calculations.
Indeed, for any of the Platonic solid measurements, unequal weightings could predictably lead to higher bounds
(attainable by a dishonest Alice by aligning Bob's LHS closer to the more highly weighted measurements), but offer no prospect of lower bounds.
The only goal for any choice of weighting (or, indeed, any choice of measurement set) is to limit the values of $S_n$ that can possibly be obtained with any cheating strategies. For an honest Alice, $S_n$ will be solely dependent upon her state's entanglement parameter $\mu$, and her efficiency. 
Thus, the only way in which measurement weightings can affect an honest Alice's capabilities is if a change in weightings changes our bounds. This is to say; if unequal weightings can lower our EPR-steering bounds, we can be certain that this is the only consequence they will effect. 

To investigate how the $n=7$ EPR-steering bound is affected when our measurement weightings are not necessarily equal, we will designate the measurement weighting for the octahedral ($n=3$) measurements as $p_3$, and for the cubic ($n=4$) measurements as $p_4$.
Our previous expressions for $D_n(i)$ used equal weightings for all measurements, so removing this restriction from Eq.~(\ref{eq:dn34}) to include a dependence upon $p_3$ and $p_4$, we obtain
\begin{equation}
D_{7}'(i)=\underset{\{A^i_{r}\}}{\max}\left[\lambda_{\max} \left(\sum_{r=1}^{n_{3}} A^i_{r}\hat{\sigma}_{r}^{\beta}\frac{p_3}{3}+\sum_{r=n_4}^{n_3+n_{4}} A^i_{r}\hat{\sigma}_{r}^{\beta}\frac{p_4}{4}\right)\right], 
\label{eq:dn34p}
\end{equation}
with $p_3+p_4=1$. Note (from this expression) that $p_3=p_4=0.5$ does not define equal measurement weightings because there are three octahedron ($n=3$) measurements and four cube ($n=4$) measurements in our set of seven. Therefore, each measurement is chosen equally often with the {\em balanced} weightings $p_3=\xfrac{3}{7}$, $p_4=\xfrac{4}{7}$.

\begin{figure}
\begin{centering}
\includegraphics[width=.9\linewidth]{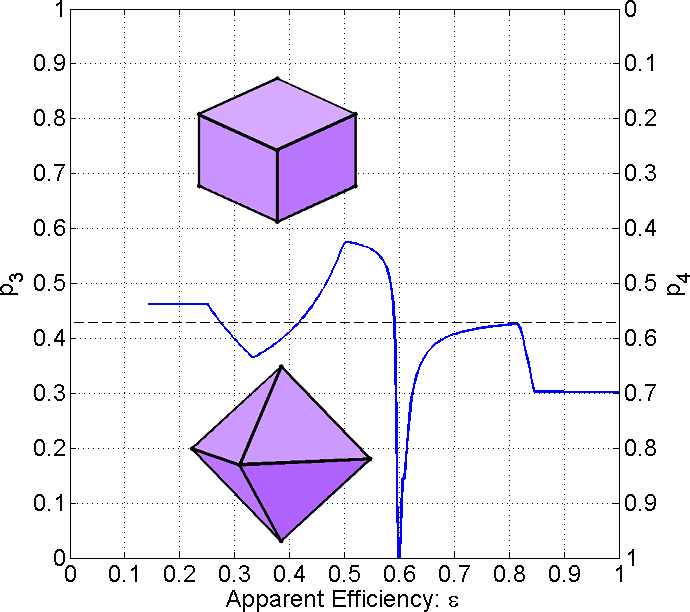}
\par\end{centering}
\caption{Optimal mixing proportions of octahedral measurements ($p_3$) and cubic measurements ($p_4$) for the variable-$p_r$ $n=7$ bounds. The dashed line indicates balanced weighting. The line stops at $\epsilon=1/7$ because EPR-steering is impossible below that point.}
\label{p3p4}
\end{figure}

When Bob chooses unbalanced weightings (which we will refer to as $p_x$ in the general case), Alice's optimal deterministic strategies will likely change. 
However, even with Eq.~(\ref{eq:dn34}) replaced by Eq.~(\ref{eq:dn34p}), Alice's optimal nondeterministic strategies are still described by Eq.~(\ref{eq:kngeo}), and we will still constrain Alice to satisfy the symmetry condition, Eq.~(\ref{eq:sym7}). Upon calculating the values of these bounds as a function of $p_x$, we find that Bob can indeed alter his $p_x$ values to lower the EPR-steering bound for almost all $\epsilon$-values. Figure~\ref{p3p4} plots the values of $p_3$ and $p_4$ that yield the lowest possible EPR-steering bounds for our $n=7$ geodesic measurements. From this figure, it is clear that EPR-steering can be more easily demonstrated by using unbalanced measurement weightings.

\begin{figure}
\begin{centering}
\includegraphics[width=.9\linewidth]{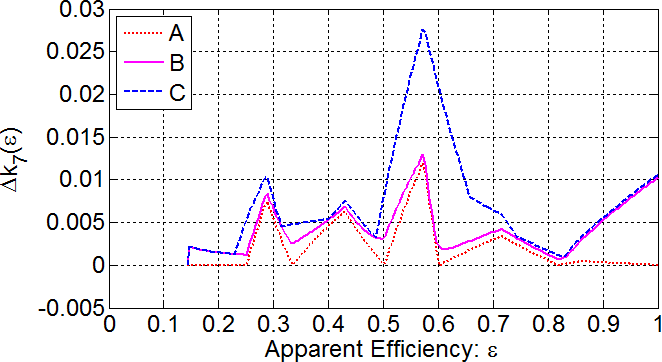}
\par\end{centering}
\caption{Numerical improvement from the asymmetric $n=7$ geodesic bounds to: (A) the symmetric bounds, (B) the variable probability bounds, and, (C) the optimal bounds.}
\label{Comparison7}
\end{figure}

Moreover, the optimal way to unbalance the correlation function is in line with the intuitive argument we used to motivate this unbalancing at the beginning of this section: to more heavily weight the measurements which give lower results in Alice's cheating strategies. Appendix A gives more detail as to how this is shown by the behaviour of Fig.~\ref{p3p4}.

However, at most $\epsilon$-values, the magnitude of the improvement we obtain in $k_7(\epsilon)$ by using the optimal weightings shown in Fig.~\ref{p3p4} is on the same scale as the difference between the two $n=7$ bounds in Fig.~\ref{fig:n7}; so it would not be very useful to plot the postselected values for these bounds. 
Instead, we have shown, in Fig.~\ref{Comparison7}, the difference between the optimally weighted bounds and the original asymmetric bounds for $n=7$ (as function ``B"). This figure also includes the difference between the asymmetric bounds and the symmetric bounds (function ``A"), so the spacing between these two functions is the degree of improvement that the optimally weighted bounds offer over the symmetric bounds. On this figure, which is approximately one-fourteenth the vertical scale of Fig.~\ref{fig:n7}, we can observe that the optimally weighted bounds offer improvement at almost every $\epsilon$-value, but given the scale upon which this change is visible, it can be said that there is not a significant improvement anywhere except near $\epsilon=1$.
The function ``C" in this graph shows the improvement gained from further types of optimisation that we discuss in the next section.


\section{Optimised Measurement Strategies} \label{sec:globaloptimal}

Our choices of measurement sets thus far have all been built upon the idea that regularly spaced measurement orientations should be of the most benefit for a rigorous test of EPR-Steering. However, in Fig.~\ref{fig:Platonic}  we saw that regularly spaced measurements for $n=4$ are definitely not optimal, and in Sec.~IV we have observed several distinct advantages that only exist for measurements that are not regularly spaced (because they combine two Platonic solids). Including 
these advantages for the $n=7$ geodesic solid gives bounds better than the Platonic bounds for $n=6$
around $\epsilon\approx0.5$. However, they are actually worse than the Platonic $n=6$ bounds for  $\epsilon\in [0.24,0.44] \cup [0.52,0.82]$, meaning that even this scheme cannot be optimal for $n=7$.

These observations motivate considering the even more general case, where we do not have two (or more) sets of measurements, 
but rather where we treat each measurement setting independently. That is, we fix only the number of settings $n$, 
and, for each $\epsilon$, optimise the $n$ directions defining the $n$ measurements, and the $n$ weightings 
defining the correlation function.  

To investigate this, we must return to our definition of $S_n$, redefining it as generally as possible.
Our use of $\{\hat{\sigma}_{r}^{\beta}\}$ already allows arbitrary measurement directions, so we need only 
define a weight for each $r$-term, which we will denote $p_r$, normalised according to
$\sum_{r=1}^{n}p_{r}=1$. 
Thus, the (non-postselected) form of $S_n$ that we consider is
\begin{equation}
S_{n}(\epsilon)=-\sum_{r=1}^{n}E_{A_{r}}\left[A_{r}\langle\hat{\sigma}_{r}^{\beta}\rangle_{A_{r}}\right]p_{r}.
\label{eq:SnFormal}
\end{equation}
%
In this scenario (just as in  Sec.~\ref{sec:IVC}), it is not actually necessary for Bob to experimentally choose measurement setting $r$ with probability $p_r$ 
in order to calculate  Eq.~(\ref{eq:SnFormal});  he can choose different settings with arbitrary frequency and 
merely weight each term appropriately in his calculation of $S_n$.

To obtain the strongest bounds for variable measurement sets, it will clearly be necessary
to employ our symmetry condition. In a form which is independent of measurement orientations (or relationships thereof),
the condition a cheating Alice must meet for her null probabilities to be independent of measurement orientation is
\begin{equation}
\sum_{i}w_{i}|A_{r}^{i}|=\epsilon,\mbox{}\forall r,
\label{eq:SymCond}
\end{equation}
where for a given deterministic strategy $i$, $A_{r}^{i}$ is the result she reports  
when Bob measured with setting $r$, and $w_i$ is the probability with which 
she chooses each strategy. Note that $|A_r^i| \in \{0,1\}$ is the efficiency for 
measurement $r$ under strategy $i$.
Alice's optimal deterministic strategies are those which attain the maximum in the expression,
\begin{equation}
D_{n}(i)=\underset{\{A^i_{r}\}}{\max}\left[ \lambda_{\max}\left(\sum_{r}^{n} A^i_{r}\hat{\sigma}_{r}^{\beta}p_{r}\right)\right].
\label{eq:dngen}
\end{equation}
This looks markedly more similar to Eq.~(\ref{eq:platdet}) than it does to Eq.~(\ref{eq:dn34}) or Eq.~(\ref{eq:dn34p}),
but its only deviation from any of these equations is that, to define it with generality, we must take the $i$ index to denote the optimal deterministic strategies for each possible permutation of null/non-null values for \textit{all} measurements.
That is, for $n$ measurements, where we now label $m_r^i=|A_r^i|\in\{1,0\}$, we must consider the optimal deterministic strategies for all $2^n$ possible values of the list of $(m_1^i,m_2^i,\ldots,m_{n-1}^i,m_n^i)$. 

Thus, to employ our generalised symmetry constraint, the maximal nondeterministic bound on $S_{n}$ 
cannot be defined by Eq. (\ref{eq:NDbound}), which is not compatible with Eq.~(\ref{eq:SymCond}),
but must be defined as
\begin{equation}
K_{n}(\epsilon)=\underset{\{w_{i}\}}{\max}\left[{\sum_{i=1}^{2^n}}w_{i}D_{n}(i)\right],
\label{eq:kngen}
\end{equation}
where, most generally, $i$ indexes the set of all possible deterministic strategies, $\{D_{n}(i)\}$.
This is because if Alice's numerically optimal nondeterministic strategies cannot be
arranged to satisfy the symmetry condition, she
will need to use some suboptimal deterministic strategies in order
to satisfy this condition (and to maintain a reasonably high value for $S_{n}$). 

While Bob's choice and implementation of $\{p_r\}$ are of no consequence to an 
honest Alice (except in their capacity to lower the EPR-steering bound), it merits brief
observation that a cheating Alice cannot attain the bounds $K_n(\epsilon)$ on $S_n(\epsilon)$ 
without knowing what $\{p_{r}\}$ will be, 
since the optimal deterministic strategies defined by Eq.~(\ref{eq:dngen}) involve $p_r$. 
(The same is true of the $p_x$ in Sec.~\ref{sec:IVC}.)
But, as described above, Bob's only priority in choosing $\{p_r\}$
is to make the EPR-steering bound as low as possible.


So, given some measurement set $\{\hat{\sigma}_r^{\beta}\}$ and set of weights $\{p_r\}$, 
we can calculate $K_n(\epsilon)$ from Eq.~(\ref{eq:kngen}).
Thus, in terms of the post-selected $S_n(\epsilon)$, as we have been using,
the EPR-steering bound is $k_n(\epsilon) = K_n(\epsilon)/\epsilon$.
Calculating which measurements and weightings minimise $k_n(\epsilon)$ requires searching simultaneously over all $\hat\sigma_r^\beta$ and $p_r$ variables.
Thus, we can only define the optimal value of $k_n(\epsilon)$ as
\begin{equation}
c_{n}(\epsilon)=\underset{\{p_{r}\}}{\min}\left[\underset{\{\sigma_{r}\}}{\min}\left(k_{n}(\epsilon)\right)\right].
\label{eq:opt}
\end{equation}
We can minimise the dimensionality of this problem by holding static the direction of the first $\hat\sigma_r^\beta$ and the plane
of the second, and defining one $p_r$ from the other $n-1$ of them (using their completeness relation), but this still leaves a search space of $3n-4$ scalar variables.
Moreover, such an optimisation is required for every different $\epsilon$ value. 
Performing such optimisations numerically does not require unreasonable amounts of computational power for moderate $n$ \footnote{The solution time for $n=4$ takes about an hour per data point in Matlab on a standard personal computer. However, every time $n$ increases by 1, the variable space requires three more dimensions. Even with efficient optimisation algorithms, our solving time still increases exponentially with $n$, almost doubling with every increase in $n$, being approximately proportional to $1.85^n$.}.
The sets $\{\hat{\sigma}_{r}^{\beta}\}$ and $\{p_{r}\}$ that 
achieve the minimum bound, $c_n(\epsilon)$, define the optimal steering experiment using Werner states, $n$ measurement settings, and an apparent efficiency of $\epsilon$.

\subsection{Optimal EPR-steering bounds for $n=4$}

We observed earlier in Fig.~\ref{fig:Platonic} that for $0.48\lesssim\epsilon\lesssim0.58$, the Platonic solid EPR-steering bound for $n=4$  (cube) was not as loss-tolerant as the $n=3$ (octahedron) bound, which would not be possible were it an optimal set of four measurements. This makes $n=4$ the obvious place to start for our optimisation.

\begin{figure}
\begin{centering}
\includegraphics[width=.9\linewidth]{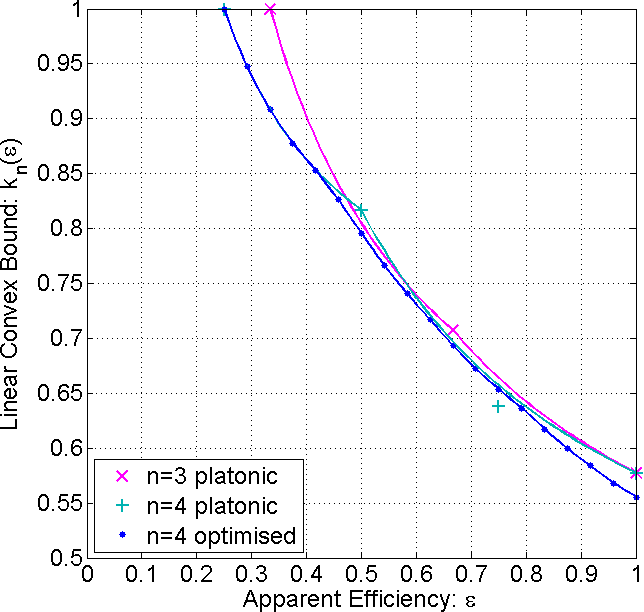}
\par\end{centering}
\caption{Comparison of optimised $n=4$ bounds with $n=3$ and $n=4$ Platonic bounds.}
\label{optbounds4}
\end{figure}

This optimisation was performed for $n=4$ at 18 different $\epsilon$ values,
with spacing $\Delta\epsilon=\xfrac{0.75}{18}\approx0.042$ between
each value. The EPR-steering bounds $c_n(\epsilon)$ yielded by each optimised
measurement strategy are shown in Fig.~\ref{optbounds4}.
For comparison, this figure also displays the Platonic solid bounds for $n=3$ and $n=4$.
One might expect us to show also the $n=3$ optimised bounds,  
but it turns out that the octahedral measurement strategy for $n=3$ is already an optimal measurement strategy for every $\epsilon$. (At least this is what we found  
after performing the optimisation for $n=3$ over a large range of $\epsilon$ values.) 
It was concluded that the same is true of the square strategy for $n=2$.

In Fig.~\ref{optbounds4}, the points on the Platonic solid curves are optimal
deterministic strategies, and the lines are the nondeterministic strategies corresponding to the optimal bounds for these measurement sets,  
as usual. But on the optimised measurement curve, the only bounds which are definitely optimal are the data points,
as these are the only $\epsilon$ values for which optimisations have been performed.
The curve connecting these points is calculated from nondeterministic mixings of these optimised bounds.
However, analysis of these data points indicates that the optimal values of $\{ p_r \}$ and $\{\hat{\sigma}_{r}^{\beta}\}$ vary
quite slowly relative to $\Delta \epsilon$, so this curve almost certainly closely approximates the intermediate optimal bounds.

As we can see, the optimal bounds for $n=4$ are lower than the $n=3$
bounds in all places, which more than fulfils our motivating requirement that
optimal $n=4$ bounds should have $c_{4}(\epsilon)\leq c_{3}(\epsilon)$
$\forall \epsilon$. Indeed, the optimised bounds are also visibly lower than the
$n=4$ Platonic solid bounds for $\epsilon \gtrsim 0.42$,
but converge with the Platonic bounds as $\epsilon\searrow0.25$. 

\begin{figure}
\begin{centering}
\includegraphics[width=.9\linewidth]{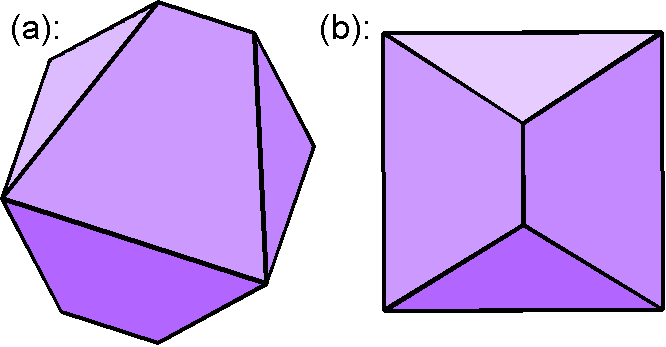}
\par\end{centering}
\caption{The solid representing the optimal measurement arrangements for $n=4$ when $\epsilon=0.5$, from two different angles. The two vertices at the top of (a) are the same two vertices in the centre of (b).}
\label{n4ep5solid}
\end{figure}

Performing the minimisation in Eq.~(\ref{eq:opt}) for $n=4$, with a large number of $\epsilon$-values, reveals that the optimal measurement strategy for $n=4$ is still to use equally weighted cubic vertices for $\epsilon \lesssim 0.42$, but as $\epsilon$ increases, the optimal measurement strategy deviates from the cube (as a seemingly continuous function of $\epsilon$), approaching the spatial configuration shown in Fig.~\ref{n4ep5solid}, which represents the optimal measurement strategy for $\epsilon =0.5$. 
The optimal values of $\{p_r\}$ at this point are such that the two measurements in the same plane---the ones that define the square visible in Fig.~\ref{n4ep5solid}(b)---have weightings of $p_r=1/3$, and the other two measurements have weightings of $p_r=1/6$.

\begin{figure}
\begin{centering}
\includegraphics[width=.9\linewidth]{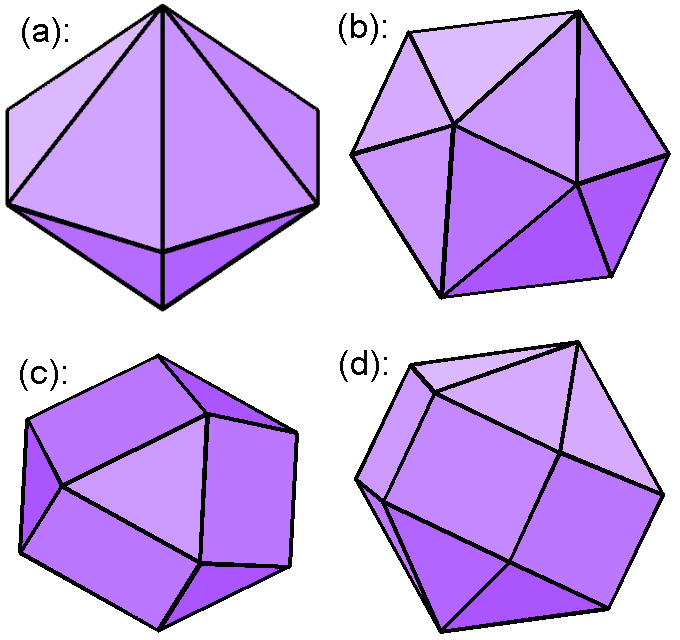}
\par\end{centering}
\caption{The solids representing the optimal measurement arrangements (when $\epsilon=1$) for (a): $n=4$, (b): $n=5$, (c): $n=6$, and (d): $n=7$. (b) and (c) can be thought of as ``top-down" compared to the perspective of (a).}
\label{OptSolids}
\end{figure}

As $\epsilon$ increases above $\epsilon=0.5$, this optimal measurement strategy undergoes another continuous transition,
and at $\epsilon= 1$, the optimal arrangement becomes that 
shown in Fig.~\ref{OptSolids}(a): three measurements almost (but not quite) equally spaced in the same plane---the optimal lengths of their edges seem to be around 1.03, 1.00, and 0.97, and this performs better than exactly equally spaced measurements---and a fourth perpendicular to them. The weightings 
associated with these measurements are $p_{r}\approx0.23$ 
for the three planar measurements, and $p_{r}\approx0.31$ 
for the extraplanar measurement. 

\subsection{Optimal EPR-steering Bounds for $n\geq 4$} 

\begin{figure}
\begin{centering}
\includegraphics[width=.9\linewidth]{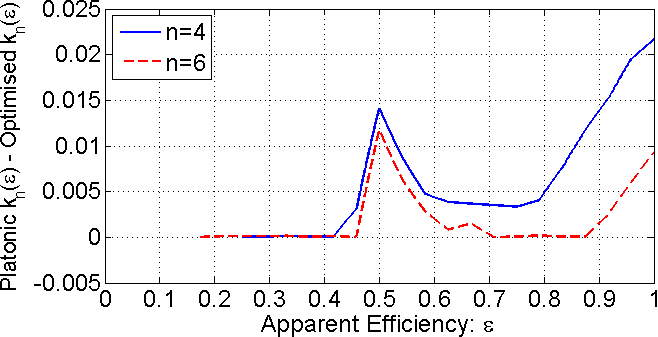}
\par\end{centering}
\caption{Numerical difference between Platonic bounds and optimised bounds for $n=4$ and $n=6$.}
\label{Comparison}
\end{figure}

\begin{figure}
\begin{centering}
\includegraphics[width=.9\linewidth]{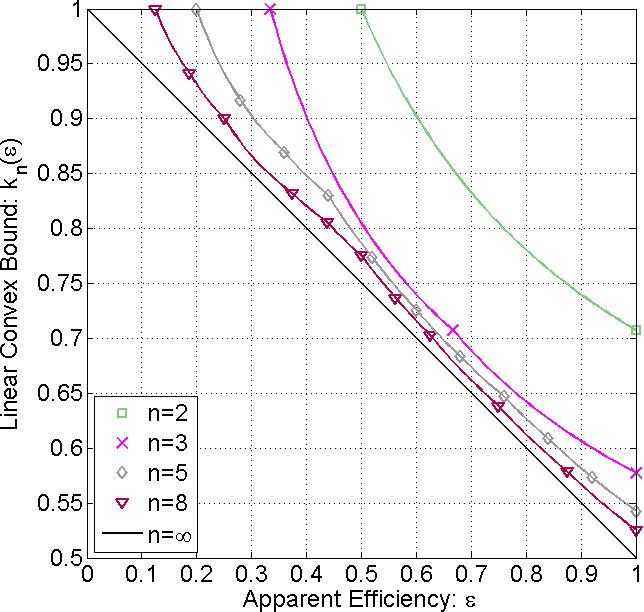}
\par\end{centering}
\caption{Optimised EPR-steering bounds for $n=2,3,5,$ and 8, with the analytical infinite measurement limit included. The $n=2$ and $n=3$ bounds are equal to the Platonic bounds.}
\label{AllOpt}
\end{figure}

Although we found the Platonic solid bound for $n=3$ to be an optimal bound,
the clear improvement of the optimal $n=4$ bounds over their Platonic
solid counterparts strongly suggests that there may be room for improvement
in the other Platonic solid measurement strategies. 
Upon calculating a series of optimal strategies for $n=5$, this suggestion becomes an insistence, since we find that the 
optimal bounds for $n=5$ are again 
better than the Platonic bounds for $n=6$ in a range near $\epsilon=0.5$ 
(we plot the $n=5$ curve later, in Fig.~\ref{AllOpt}).  Calculating optimal measurement strategies for $n=6$ gives bounds that are, as expected, equal to the Platonic bounds in some places, but slightly better in most.
In Fig.~\ref{Comparison}, we have plotted the quantitative improvement that the optimised bounds offer over the Platonic bounds for $n=4$ and $n=6$ (more visibly displayed than the form of Fig.~\ref{optbounds4} allows).

The maxima and minima in Fig.~\ref{Comparison} are indicative of the advantages that optimised measurement strategies offer over Platonic measurements, so we explain in Appendix B what causes them to occur.
At each point, it seems that the most beneficial measurement sets should generally be reasonably close to being regularly spaced, but not quite. The most beneficial $\{p_r\}$ sets merely augment these properties, with most $p_r$ being close to equal, but slightly higher for measurements that are the most outlying.

Based on an exploration of optimised strategies for $4\leq  n\leq10$ (though less comprehensively for $n=9$ and $n=10$), similar behaviours seem to be generally applicable to the optimal strategies for any $n$.
Indeed, the optimal measurement arrangements for $n=5,6,$ and 7 have obvious traits in common with those for $n=4$.
If we define the vertices of a solid from our optimised measurement orientations, we obtain solids for $n=5$ and $n=6$ that have almost the same arrangement of three equatorial vertex pairs that $n=4$ elicits. For $n=5$ and $n=6$, the only substantial difference from the $n=4$ case is that the single vertex at the top of that figure is replaced by a pair of vertices for $n=5$, and a (scalene, but nearly equilateral) triangle of vertices for $n=6$.  

\begin{figure}
\begin{centering}
\includegraphics[width=.9\linewidth]{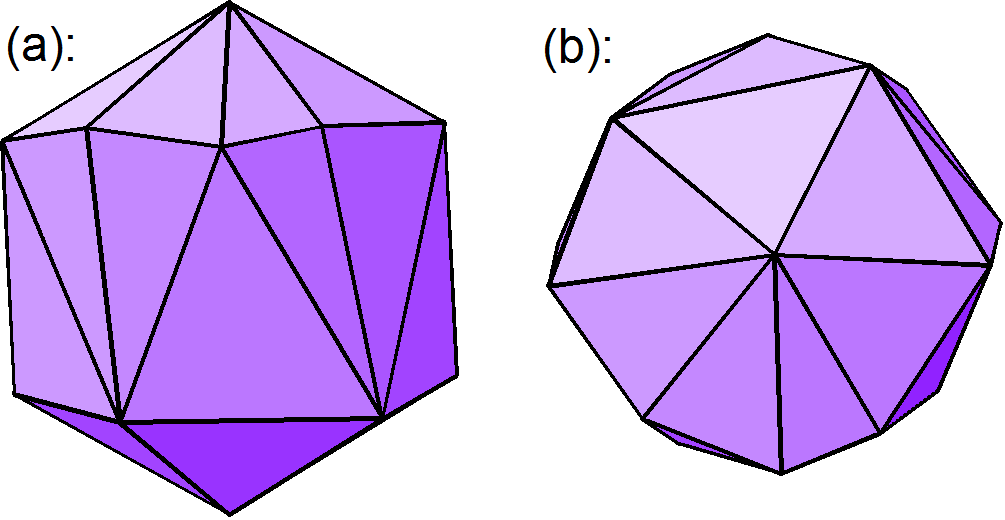}
\par\end{centering}
\caption{The solid representing the optimal measurement arrangements for $n=8$ when $\epsilon=1$, from two different angles. The vertex at the top of (a) is the same vertex as in the centre of (b).}
\label{n8solid}
\end{figure}

This property is made as visible as possible in Figs.~\ref{OptSolids}(b) and \ref{OptSolids}(c), with their three planar vertex pairs being the six outermost vertices visible on both of those images. The optimal solid for $n=7$, on the other hand, breaks with this pattern, but still shows a noticeable similarity to the $n=4$ shape. Shown in Fig.~\ref{OptSolids}(d), this solid has the same top-down profile as the $n=4$ solid,
centred on a ``top-bottom" vertex pair. Unlike $n=4$, the remaining vertices are not arranged in a single plane, but are arranged in two parallel planes with three vertex pairs defining each one---which is the source of the similarity between our $n=4$ and $n=7$ shapes.
The optimal $n=8$ solid shown in Fig.~\ref{n8solid} does not bear an immediate resemblance to any of the other optimal solids in Figs.~\ref{n4ep5solid} or \ref{OptSolids},
but can easily be seen to approximate two parallel planes of vertices with and another vertex orthogonal to them.
The solids shown in Figs.~\ref{OptSolids} and \ref{n8solid} were all generated from $\epsilon=1$ optimisations, and do change slightly with $\epsilon$, but retain the same general arrangements at all points.

In addition to this, the optimal EPR-steering bounds for $n=4,5,6,7,$ and 8 
all seem to adopt the same general behaviour that we have observed in our analysis of the above bounds.
If we return to Fig.~\ref{Comparison7}, we can see that the improvement of the optimised $n=7$ bounds over the other examples does follow a similar pattern to that observed in Fig.~\ref{Comparison} for $n=4$ and 6. Around $\epsilon\approx 0.3$ and $\epsilon\approx 0.75$, our optimised $n=7$ bounds offer little improvement upon their more regularly spaced counterparts, for the same reasons described above.
 However, Fig.~\ref{Comparison7} shows that for $n=7$ (at least), the improvements of the optimised bounds at $\epsilon\approx1$ are largely due to the advantage of unequal measurement weightings.

Returning to Fig.~\ref{Comparison}, a final trend to discuss is that the improvement in $n=4$ bounds, at all points, exceeds the improvement in $n=6$ bounds. Seeing as the Platonic $n=4$ bounds were the only ones to be outperformed by another Platonic solid at any point, this is not surprising.
However, perhaps a better framing of the reasons for this can be seen in the tendency of higher $n$-values to yield bounds ever closer to the infinite measurement limit---the lowest possible values that EPR-steering bounds can take, regardless of measurement number---analytically calculated in Ref.~\cite{key-3}. 
This limit can be expressed as a diagonal line on our graphs, and is shown in Fig.~\ref{AllOpt}.
As $n$ increases, the Platonic bounds (in Fig.~\ref{fig:Platonic}) approach this diagonal, but with every step towards it being smaller than the last (with respect to their increases in $n$).
We would expect that optimised EPR-steering bounds should also approach this limit in a similarly asymptotic manner, albeit more swiftly than sub-optimal bounds. Therefore, it should be reasonable to expect that the closer a Platonic bound is to the $n=\infty$ line, the smaller the advantage conferred by optimising it, just as with the advantage conferred by increasing measurement number.
%

As expected, we find that the optimised EPR-steering bounds do approach the $n=\infty$ bound more quickly (with respect to $n$) than the Platonic bounds do. The optimised bounds for $n=2,3,5,$ and 8 are shown in Fig.~\ref{AllOpt}, and at almost every $\epsilon$-value, the optimised $n=8$ bounds seen here are actually closer to the diagonal than the Platonic $n=10$ bounds are (especially around $\epsilon=0.5$, where the $n=10$ bounds are inferior to every optimised bound with $n>4$). Indeed, the proximity of the bounds in Fig.~\ref{AllOpt} to the diagonal limit shows that with $n=8$, these measurement strategies are considerably loss-tolerant, and have very little room for improvement. However, any optimised strategy with $n>8$ is guaranteed to be at least as loss-tolerant, and at least as close to the diagonal as the best bounds in Fig.~\ref{AllOpt} (and will necessarily be incrementally closer for \emph{at least} some range near $\epsilon=\xfrac{1}{n}$).

In Fig.~\ref{AllOpt}, this trend is easily observed, but here we can also see the relevance of regularly spaced measurements being close to optimal around $\epsilon \approx0.3$ and $\epsilon \approx0.75$: Around these places, the Platonic bounds (for $n=6$ and 10, at least) were already reasonably close to their graph's diagonal. Thus, it stands to reason that these would be $\epsilon$ values where the possible advantages of any other measurement strategies would generally be most limited. This also offers insight as to why the greatest advantages for our optimised bounds were around $\epsilon\approx0.5$ and $\epsilon\approx 1$. Such behaviour is reassuring to see in optimised bounds, since it is reasonable that only with optimal bounds can we see higher $n$-values necessarily leading to bounds that are incrementially closer to a diagonal line each time.

\section{Conclusion}

In our consideration of EPR-steering tests for two-qubit Werner states, we have confirmed our earlier conclusion \cite{Evans}
that the detection loophole in these tests can be closed without necessarily placing any particularly demanding experimental constraints upon one's detection efficiency. This can be accomplished by employing a large number $n$ of measurements in each test. However, we have also shown, contrary to the assumptions of previous experiments, that measurement sets based upon Platonic solids are, in general, suboptimal. 
  
Of course, Platonic solids are suboptimal in that they are restricted to $n\leq 10$, but this limit can be overcome by  combining Platonic solids to make geodesic solids (which can be defined for arbitrarily large $n$ if desired). The more interesting point is that Platonic solids are demonstrably suboptimal even for $n$ as small as $4$.  
Specifically for some values of Alice's efficiency, there are Werner states which do not violate the EPR-steering 
inequality for the $n=4$ Platonic solid, when we know that EPR-steering can be demonstrated even with $n=3$.
  
Considering geodesic solids and how to test Alice's steering ability most rigorously pointed the way to defining 
 the optimal steering tests for any $n$, even those for which there exist no Platonic solid or geodesic solid.
 This means that more measurements can always yield more loss-tolerant tests of EPR-steering. 
More importantly, it means that even with $n$ relatively small, tests of EPR-steering can be much more loss-tolerant than with Platonic solids, or any other merely intuitive strategies.

We calculated and explored the optimal measurement strategies for measurement numbers of $n=3,4,5,6,7,$ and 8, but were
prevented from easily exploring the optimal strategies for $n\geq 9$ by the computational demands of their numerical derivations. For this reason, we should conclude that geodesic measurement sets may be a more practical alternative than truly optimal measurements for loss-tolerant experimental tests of EPR-steering for large $n$. Optimising the EPR-steering inequalities for geodesic measurements do require numerical minimisation, but the number of parameters scales only 
logarithmically with the number of settings $n$. This is significantly less demanding than generating a fully optimal measurement strategy and inequality for $n$ settings, which has $3n-4$ free parameters.
We note that a recent paper \cite{QuantEPR} has suggested an alternate method for demonstrating steering 
with large numbers of measurements, by using random bases, although without consideration of inefficiency or loss.

Finally, we note that further work would be required to turn the EPR-steering inequalities we have derived here into 
truly experimentally applicable inequalities. There are two reasons for this. First, we have assumed that Bob's detectors 
are completely characterised, with no unknown systematic errors. Second, we have allowed Bob to place 
restrictions on Alice's reported results (that the frequencies of nulls are independent of his setting) 
which cannot be exactly verified from any finite data set. A completely rigorous experimental test would have to 
include the (very small) increase in the ability of an untrusted Alice to cheat by exploiting the imperfections 
of Bob's measurement apparatus, and any allowed deviation of her null-rates from the average.

\subsection*{Acknowledgements}
We thank Cyril Branciard for initially pointing out that the cubic arrangement could not possibly 
be optimal for $n=4$. This work was supported by ARC Centre of Excellence Grant No. CE110001027.

\appendix
\section{BEHAVIOUR OF FIG.~\ref{p3p4}}

The behaviour of Fig.~\ref{p3p4} is clearly meaningful, though its implications are not immediately obvious. It supports the general hypothesis that optimal weightings entail lower weights for measurements with higher results in Alice's optimal cheating strategies.
For example, the peak in $p_3$ at $\epsilon=0.5$ in Fig.~\ref{p3p4} occurs in a region where the average results of the $n_3$ measurements are lower than the $n_4$ measurements. At this point, Alice's optimal mixing of deterministic strategies is the same as at $\epsilon=0.5$ in Fig.~\ref{fig:weighting7}, this being a 50$\%$ mix of the (1,2) and (2,2) strategies.
The optimal configurations of non-null measurements in the (2,2) strategy are such that the optimal LHS orientations shown on the (2,2) solid in Fig.~\ref{fig:weighting7} are always the optimal orientations for the (2,2) strategy, regardless of $p_3$, and these give slightly higher results for the $n_3$ measurements, on average. The (1,2) strategy is not as symmetric as the (2,2) strategy, and its optimal LHS orientations do change with $p_3$, but they still give higher average results for the $n_4$ measurements with any $p_x$ such that $p_3\leq 1/\sqrt{3}\approx 0.58$ (the difference between these average $n_3$ and $n_4$ results being much greater than the same difference in the (2,2) strategy). Above that, the (1,2) ensembles start giving higher $n_3$ results, on average. Thus, the results of Alice's optimal cheating strategy at $\epsilon=0.5$ are lowest when Bob chooses $p_3 \approx 1/\sqrt{3}\approx 0.58$.





The particularly obvious discontinuity at $\epsilon=0.6$ in Fig.~\ref{p3p4} is also caused (in part) by the marked symmetry of the (2,2) strategy, but in conjunction with the (1,4) strategy. Just as for the (2,2) strategy, the optimal LHS orientation for the (1,4) strategy does not change no matter how $p_x$ changes. Unlike (2,2), the (1,4) strategy yields higher average results for the $n_4$ measurements than the $n_3$ measurements. However, the symmetry condition restricts Alice's mixture of these two strategies at $\epsilon=0.6$ to weight the (2,2) strategy four times more heavily than the (1,4) strategy (as can be seen in Fig.~\ref{fig:weighting7}). The difference between the average $n_3$ and $n_4$ results in (2,2) is not greater than that difference in (1,4), but this difference in (2,2) is greater than a quarter of the difference in (1,4). Since the symmetry condition requires Alice to use an 80:20 ratio of these two strategies, any increase in $p_4$ will thus lower $k_7(0.6)$.
%
%
Because $p_x$ does not affect the optimal LHS orientations for (2,2) or (1,4), the optimal value of $p_4$ at $\epsilon=0.6$ is therefore $p_4=1$.
This kind of collapse is only possible when the optimal mixture is composed only of strategies with the symmetric characteristics of the (2,2) and (1,4) strategies. This mixture itself is only the optimal one because the differences between the average $n_3$ and $n_4$ results, in both the (2,2) and (1,4) strategies, is small enough that even when $p_4=1$, there simply does not happen to be any other possible mixture of strategies that can attain as high a value of $S_7(\epsilon)$ while still maintaining $\epsilon=0.6$. Indeed, we can see in Fig.~\ref{Comparison7} that the improvement at $\epsilon=0.6$ is only about $\Delta k_7(0.6)\approx 0.002$.
Thus, this anomaly provides a good example of how strategically unbalanced weightings allow us to optimally utilise a chosen set of geodesic measurements.

\section{BEHAVIOUR OF FIG.~\ref{Comparison}}

\paragraph{No advantage at $\epsilon \lesssim 0.4$.}
At $\epsilon = 1/n$, all measurement strategies necessarily yield the same bounds, and for other efficiencies close to $\epsilon=1/n$, a cheating Alice can easily choose nulls for most of her measurement results, and select states for Bob that are equidistant between the non-null measurements. Thus, for these $\epsilon$ values, being as far apart as possible (i.e., regularly spaced) is of most importance in a measurement strategy, which is why the Platonic measurements are optimal or quite nearly optimal in this range.

\paragraph{Little or no advantage at $\epsilon\approx 0.75$.}
A similar principle applies for the minima to the right of the central peak in Fig.~\ref{Comparison}, when Alice must choose most of her results to be non-null. With non-regularly spaced measurements, Alice can often compose her deterministic strategies for $\epsilon_m\approx 0.75$ to have non-null arrangements that are more closely spaced than Platonic measurements allow (in a non-regular set of orientations, it's easy to find one or two that are more isolated from the others, whereas in a regular set, this is impossible). The symmetry condition curbs this ability somewhat since it requires each measurement to have the same average non-null probability, but when Alice has several $m$-values between $\epsilon\approx 0.5$ and $\epsilon=1$ (i.e., when $n$ is large), it becomes easier for her to mix these closely spaced high-$m$ ($\epsilon_m>0.75$) strategies with low-$m$ strategies that give higher expectation values for the outlying measurements. This is why the Platonic strategies are close to optimal around $\epsilon\approx 0.75$, and moreso for $n=6$ than $n=4$.

\paragraph{Large advantage at $\epsilon \approx 0.5$.}
In this regime, Alice's optimal strategy is to choose roughly the same number of nulls and non-nulls in each deterministic strategy 
\footnote{Alice's other option would be to mix high-$\epsilon$ and low-$\epsilon$ strategies. 
However, this would combine the weaknesses of the low-$\epsilon$ and high-$\epsilon$ strategies without optimally employing their strengths. 
Alice's most effective cheating strategies at low $\epsilon$ are those which 
maximise Bob's results for a minority measurements by disregarding his results for the majority (which are assigned nulls). At high $\epsilon$, Alice's most effective strategies are those which maximise Bob's results for as many measurements as possible, with her ability to do so being limited by how many measurements she can afford to assign nulls for (and therefore not care about their values). Choosing roughly the same number of nulls and non-nulls in each deterministic strategy allows 
her to most effectively prioritise the maximisation of half of Bob's measurements over the other half, which she can afford to not care about (and submit nulls for) in each deterministic strategy.}.
To do this, Alice would need to find closely-spaced configurations of $m\approx\xfrac{n}{2}$ measurements to be non-null, 
and must find such configurations in as many directions as possible. This task is trivial with Platonic measurements, as their symmetry groups are such that a configuration of $m$ nearest-neighbour measurements is the same configuration for \textit{any} $m$ nearest neighbours.
Therefore, choosing a set of measurements that are not regularly spaced offers an advantage in this region.
For the $n=4$ solid in Fig.~\ref{n4ep5solid}, for example, every measurement pair is farther apart than the measurement pairs in the cubic arrangement, with the exception of the pair of lowest-weighted measurements. Thus, there is only one pair of measurements that can offer deterministic strategies lower than the cube's, and
they have very low weightings.
It is in this way that non-regularity of the optimal measurement sets is easily used to outperform Platonic measurement sets. 

\paragraph{Large advantage at $\epsilon=1$.}
Alice's strategies are most strongly restricted at $\epsilon=1$, where a cheating Alice's optimal strategy is to align
Bob's state with the spatial average of all of his measurement axes, with a suitable choice of sign. For the Platonic solids, this means choosing ensembles that are either face-centred or vertex-centred all about the Platonic solid.
However, optimised measurement strategies for $\epsilon\approx 1$ tend to have (up to $n=8$, at least) most of their measurements
defining a single plane (or two parallel planes) of vertices, and the rest clustered near the directions perpendicular to this plane.
The benefit this offers is that the spatial average(s) of all of these measurements will be farther from most of them than the Platonic averages are from their constituent measurements. 

\end{document}